\documentclass[showpacs,prl, reprint,superscriptaddress,preprintnumbers,longbibliography,twocolumn]{revtex4-1}

\usepackage{amssymb}
\usepackage{amsmath}
\usepackage{graphicx}
\usepackage{multirow}
\usepackage{hyperref}
\usepackage{xcolor}
\hypersetup{colorlinks,
	linkcolor={blue!50!black},
	citecolor={blue!50!black},
	urlcolor={blue!80!black}
}

\usepackage{color}

\begin{document}
\DeclareGraphicsExtensions{.pdf}

\title{Crossover of Charge Fluctuations across the Strange Metal Phase Diagram}
\author {Ali Husain\footnote{ahusain6@illinois.edu}}
\affiliation{Department of Physics and Materials Research Laboratory, University of Illinois at Urbana–-Champaign, Urbana, IL 61801, USA}
\author {Matteo Mitrano}
\affiliation{Department of Physics and Materials Research Laboratory, University of Illinois at Urbana–-Champaign, Urbana, IL 61801, USA}
\author {Melinda S. Rak}
\affiliation{Department of Physics and Materials Research Laboratory, University of Illinois at Urbana–-Champaign, Urbana, IL 61801, USA}
\author {Samantha Rubeck}
\affiliation{Department of Physics and Materials Research Laboratory, University of Illinois at Urbana–-Champaign, Urbana, IL 61801, USA}
\author {Bruno Uchoa}
\affiliation{Department of Physics and Astronomy, University of Oklahoma, Norman, OK 73069, USA}
\author {Katia March}
\affiliation{Department of Physics, Arizona State University, Tempe, AZ 85287, USA}
\author {Christian Dwyer}
\affiliation{Department of Physics, Arizona State University, Tempe, AZ 85287, USA}
\author {John Schneeloch}
\affiliation{Condensed Matter Physics and Materials Science Department, Brookhaven National Laboratory, Upton, NY 11973, USA}
\author {Ruidan Zhong}
\affiliation{Condensed Matter Physics and Materials Science Department, Brookhaven National Laboratory, Upton, NY 11973, USA}
\author {Genda D. Gu}
\affiliation{Condensed Matter Physics and Materials Science Department, Brookhaven National Laboratory, Upton, NY 11973, USA}
\author {Peter Abbamonte\footnote{abbamonte@mrl.illinois.edu}}
\affiliation{Department of Physics and Materials Research Laboratory, University of Illinois at Urbana–-Champaign, Urbana, IL 61801, USA}

\date{\today}

\begin{abstract}

A normal metal exhibits a valence plasmon, which is a sound wave in its conduction electron density. The mysterious strange metal is characterized by non-Boltzmann transport and violates most fundamental Fermi liquid scaling laws. A fundamental question is: Do strange metals have plasmons? Using momentum-resolved inelastic electron scattering (M-EELS) we recently showed that, rather than a plasmon, optimally-doped Bi$_{2.1}$Sr$_{1.9}$Ca$_{1.0}$Cu$_{2.0}$O$_{8+x}$ (Bi-2212) exhibits a featureless, temperature-independent continuum with a power-law form over most energy and momentum scales [M. Mitrano, PNAS {\bf 115}, 5392-5396 (2018)]. Here, we show that this continuum is present throughout the fan-shaped, strange metal region of the phase diagram. Outside this region, dramatic changes in spectral weight are observed: In underdoped samples, spectral weight up to 0.5 eV is enhanced at low temperature, biasing the system towards a charge order instability. The situation is reversed in the overdoped case, where spectral weight is strongly suppressed at low temperature, increasing quasiparticle coherence in this regime. Optimal doping corresponds to the boundary between these two opposite behaviors at which the response is temperature-independent. 
Our study suggests that plasmons do not exist as well-defined excitations in Bi-2212, and that a featureless continuum is a defining property of the strange metal, which is connected to a peculiar crossover where the spectral weight change undergoes a sign reversal.

\end{abstract}

\maketitle

\section{Introduction}

The enigmatic and poorly understood strange metal has been found within the phase diagrams of many strongly-correlated systems, including transition metal oxides, heavy fermion materials, organic molecular solids, and iron-pnictide superconductors \cite{norman11,hussey04,hussey06,stewart01, stewart06,dressel11,kasahara10}. This phase is characterized by violation of fundamental Fermi liquid scaling laws, and by its close proximity to other exotic phases such as unconventional superconductivity, charge or spin density waves, and nematicity \cite{stewart01,dressel11,norman11,ginsberg98}. For example, the prototypical copper-oxide strange metals, which are also high temperature superconductors, exhibit a resistivity that is linear in temperature and exceeds the Mott-Ioffe-Regel limit \cite{keimer15,hussey04}, an optical conductivity exhibiting an anomalous power law dependence on frequency \cite{basov05,vdmarel2003}, a magnetoresistance that is linear in field, violating Kohler's rule \cite{giraldo2018}, a quasiparticle decay rate that scales linearly with energy \cite{valla2000}, and an NMR spin relaxation rate that violates the Korringa law \cite{berthier1993}. No generally accepted theory of matter can explain these properties, which appear to be incompatible with fundamental assumptions of Boltzmann transport theory. The strange metal has thus become one of the great unsolved problems in condensed matter physics. 

The quasiparticle dynamics of strange metals have been studied extensively with angle-resolved photoemission (ARPES) and scanning tunneling microscopy (STM) techniques, which directly measure the one-electron spectral function \cite{damascelli03,reber15,bok16,vishik2010,chatterjee2011,fujita2012}. However, little is known about the two-particle charge response, which directly reveals the strongly correlated nature of this phase \cite{zaanen2006}. 

The fundamental charge collective mode of an ordinary metal is its plasmon, which is essentially a sound wave in its valence electron density \cite{nozieres99}. On the other hand, we recently showed that in optimally doped  Bi$_{2.1}$Sr$_{1.9}$Ca$_{1.0}$Cu$_{2.0}$O$_{8+x}$ (Bi-2212), a known strange metal, the plasmon is overdamped and rapidly decays into a momentum-, energy-, and temperature-independent continuum extending up to an energy of $\sim 1$ eV \cite{mitrano18}. Initial measurements away from optimal doping, on heavily overdoped samples, showed a surprising depletion of spectral weight below 0.5 eV at low temperature \cite{mitrano18}. 
This energy scale is more than $20\times$ larger than the temperature scale on which these changes take place, suggesting strong interactions are at play \cite{basov2011}. It is therefore crucial to map out the exact region of the phase diagram where a featureless continuum is present, so its connection to the strange metal and other neighboring phases can be established.
 
Here, we present a study of the density fluctuations across the doping-temperature phase diagram of the strange metal Bi-2212 using momentum-resolved electron energy-loss spectroscopy (M-EELS) \cite{vig17}. This technique measures the surface dynamic charge response of a material, $\chi(\mathbf{q},\omega)$, and directly reveals the charged bosonic collective modes of the system \cite{vig17}. Note that, while generally regarded as a surface technique, the probe depth of M-EELS is given by the inverse of the in-plane momentum transfer, $q^{-1}$, making it somewhat more bulk sensitive than single-particle spectroscopies like ARPES and STM \cite{vig17}. In this study we focus on the energy regime 0.1 eV $< \omega < 2$ eV, which is relevant to the high-temperature normal state out of which superconductivity and other instabilities form. 

\section{Experiment}





Single crystals of Bi-2212 were grown using floating zone methods described previously \cite{wen2007}. The current study was done on underdoped crystals with $T_c$=50 K (UD50K) and 70 K (UD70K), optimally doped crystals with $T_c$=91 K (OP91K), and overdoped crystals with $T_c$=50 K (OD50K).

Measurements of the charge fluctuation spectra were performed using meV-resolution, momentum-resolved electron energy-loss spectroscopy (M-EELS) \cite{vig17}. M-EELS is a variant of surface HR-EELS \cite{ibachMillsBook} in which the momentum transfer of the probe electron is measured with both high resolution and accuracy \cite{vig17}. Measurements were performed on cleaved single crystals of Bi-2212 at 50 eV incoming electron beam energy and 4 meV energy resolution at a fixed out-of-plane momentum transfer, $q_z=4.10 \textrm{ \AA}^{-1}$. We use Miller indices, $(H,K)$, to designate an in-plane momentum transfer $q=(2\pi H/a,2\pi K/a)$, where $a=3.81 \textrm{ \AA}$ is the tetragonal Cu-Cu lattice parameter \cite{damascelli03}. Unless otherwise specified, all momenta are along the (1,-1) crystallographic direction, i.e., perpendicular to the structural supermodulation in this material \cite{damascelli03}. The sample orientation matrix was determined {\it in situ} using the (0,0) specular and (1,0) Bragg reflections, and verified by observing the (1,-1) reflection, establishing a quantitative relationship between the momentum, $q$, and the diffractometer angles. M-EELS spectra were taken from 0 to 2 eV energy loss and binned into 30 meV intervals for improved statistics.

M-EELS measures the dynamic structure factor of a surface, $S(\mathbf{q},\omega)$, which is proportional to the surface dynamic charge susceptibility, $\chi''(\mathbf{q},\omega)$, by the fluctuation-dissipation theorem \cite{ibachMillsBook,kogar2014,vig17}. $\chi''$ was determined from the raw data by dividing the M-EELS matrix elements, which depend on the momentum transfer \cite{vig17}, and antisymmetrizing to remove the Bose factor \cite{vig17,mitrano18}. The data were placed on an absolute scale by performing the partial $f$-sum rule integral,

\begin{equation}
\int_0^{2 \, \textrm{eV}} \omega \, \chi''(\mathbf{q},\omega)  \, d\omega = - \pi \frac{q^2}{2m} N_{\textrm{eff}},
\end{equation}

\noindent
where $N_{\textrm{eff}}$ was determined by integrating the $q=0$ ellipsometry data from Ref. \cite{levallois16} over the same energy interval at the corresponding values of temperature and doping. 

\section{M-EELS Data}

\begin{figure}
	\includegraphics[width=8.4cm]{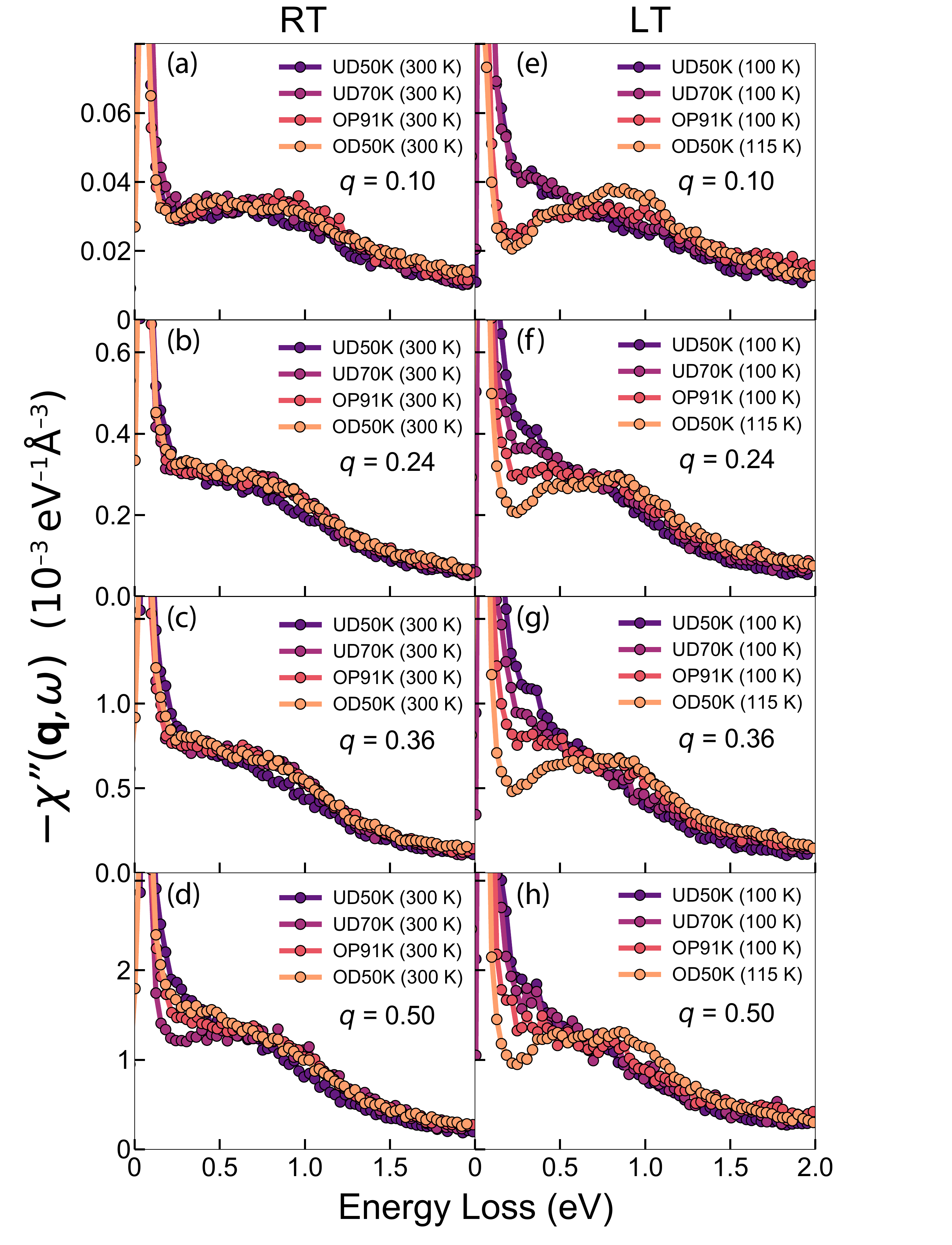}
	\caption{$\chi^{\prime\prime}(\mathbf{q},\omega)$ at room temperature (300 K) (a-d) and low temperature (either 100 K or 115 K) (e-h) for all four dopings studied. No significant $q$-dependence is seen for any doping or temperature. At 300 K the spectra vary only slightly with doping, compared to the dramatic doping dependence below  $\sim$0.5 eV at low temperature.}
	\label{300Kvs100K}
\end{figure}

Figure \ref{300Kvs100K} shows the M-EELS spectra for a selection of momenta at room temperature (300 K) and at low temperature (100 K or 115 K, depending upon the doping). The intensity rise below 0.1 eV in all spectra is due to the well-known Bi-2212 optical phonons \cite{vig17,mills94,qin10}. Looking at the 300 K data, the lowest momentum (Fig. 1(a)) shows a highly damped plasmon reported previously \cite{mitrano18},which appears in the spectrum as a local maximum at $\omega \sim 1$ eV energy loss. This peak has very similar shape in all four dopings studied, and has the same energy and width as the plasmon observed in transmission EELS measurements at the same in-plane momentum \cite{vig17,mitrano18,nucker1991}. Previous studies indicate that the optical properties of the cuprates are dominated below $\sim 1$ eV by excitations in the CuO$_2$ planes, but that interband transitions, potentially involving the BiO layers, contribute above 1.5 eV and may create features around 2.3 eV and 3.8 eV \cite{quijada1999, bozovic1990,giannetti2011}. We therefore expect the M-EELS continuum principally arises from the CuO$_2$ planes but may have other contributions at high energy. 

At larger momenta (Fig. 1 (b)-(d)) the plasmon is no longer present, by which we mean that a local maximum is no longer observed in the spectra. Instead, the plasmon decays into an energy-independent continuum, as reported previously \cite{mitrano18}. As shown in Appendix B, this continuum appears also in transmission EELS experiments, and is thus a bulk property of Bi-2212. This continuum is not a subtle effect; it saturates the $f$-sum rule and is the primary feature of the charge response of Bi-2212. As reported in Ref. \cite{mitrano18}, the M-EELS spectra exhibit very little $q$-dependence at momenta greater than 0.16 r.l.u., which implies that charge excitations barely propagate in this material. Additionally, Fig. 1 (a)-(d) shows that, at $T=300$ K, the continuum also has little doping dependence across the composition range.

Surprisingly, the spectra become dramatically doping-dependent at low temperature (Fig. 1(e)-(h)), which is the main finding of this work. In overdoped samples, the spectral weight below 0.5 eV is greatly suppressed at low temperature \cite{mitrano18}. By contrast, in underdoped materials, the weight in this energy range at low temperature is enhanced. The 0.5 eV energy scale of this spectral weight rearrangement at all compositions is more than an order of magnitude larger than the temperature scale on which it occurs, indicating that the effect arises from strong electron-electron interactions. The enhancement in the low-energy susceptibility at low temperature suggests that underdoped Bi-2212 should have a tendency to form charge density waves (CDW), which may be connected to recent observations from resonant x-ray scattering \cite{dasilveneto2014,comin2016} (note that we have not seen evidence for a true CDW in Bi-2212 with M-EELS, though further study may yet uncover such effects). All spectra are doping- and temperature-independent above 1 eV where the system exhibits a universal $\sim 1/\omega^2$ form.


\begin{figure}
	\includegraphics[width=8.0cm]{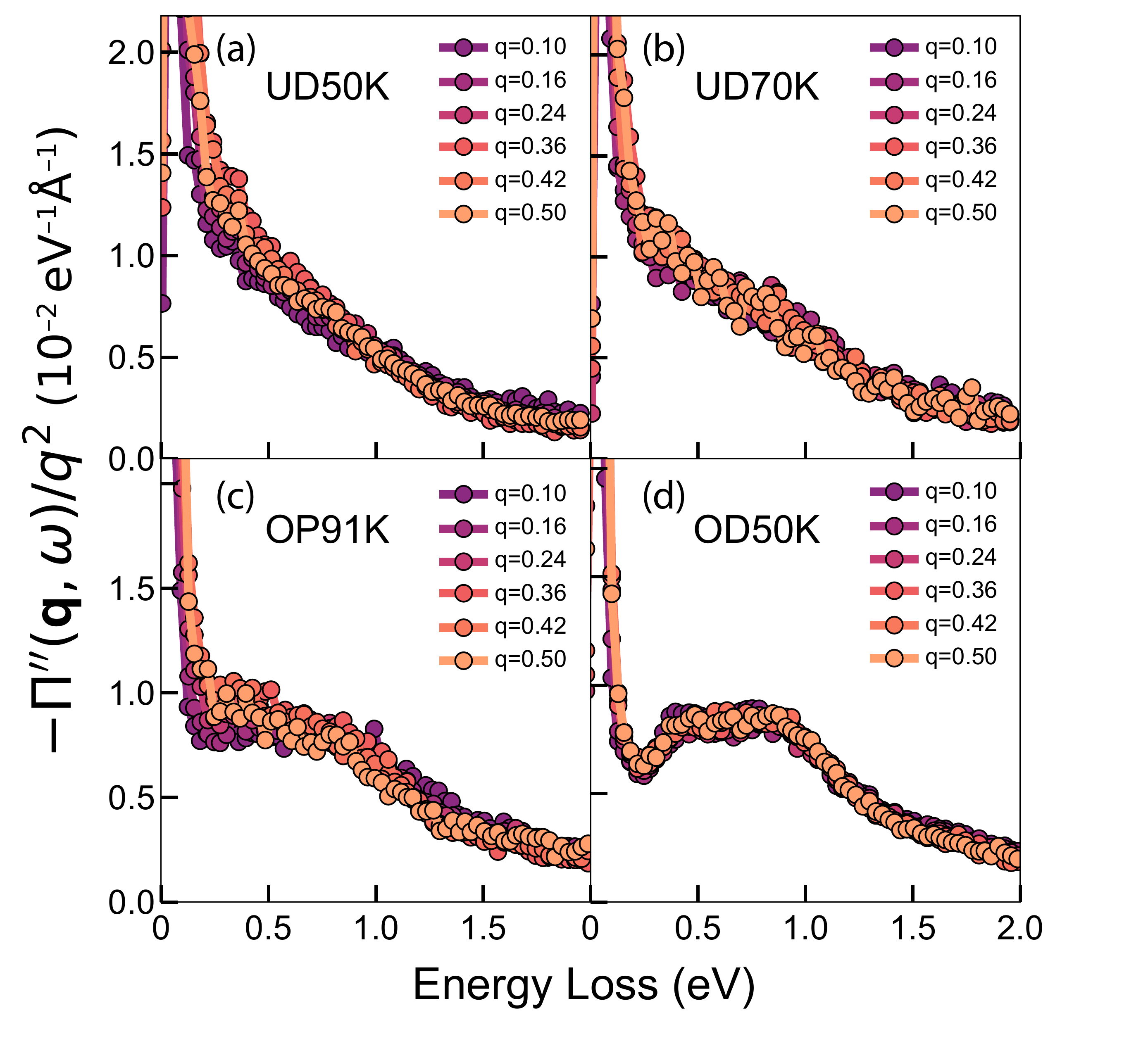}
	\caption{Collapse of  $\Pi''(\mathbf{q},\omega)$ for all momenta for each of the four doping values studied, achieved simply by dividing by $q^2$. \textcolor{black}{The spectra were taken at $T=100$K except for the OD50K sample, which was measured at 115 K.} Apart from the $q^2$ scaling of the magnitude, which is required by the $f$-sum rule, the spectra are essentially momentum-independent.}
	\label{collapse}
\end{figure}

\textcolor{black}{\section{Properties of the polarizability}}

The continuum is essentially momentum-independent at all doping values for $q>0.16$ r.l.u., where there is hardly any discernible difference between M-EELS spectra at different $q$ values (Fig. 1). \textcolor{black}{This implies that the susceptibility, $\chi''(\mathbf{q},\omega)$, is constant over $90\%$ of the Brillouin zone, apart from an overall $q^2$ dependence required by the $f$-sum rule (Eq. 1). Generally speaking, the momentum dependence of the susceptibility $\chi(\mathbf{q},\omega)$ is determined by two separate effects: the intrinsic polarizability of the system, $\Pi(\mathbf{q},\omega)$, and the momentum dependence of the Coulomb interaction itself, $V(\mathbf{q})$. It is therefore  important to determine whether the deviation from a constant in the remaining portion of the Brillouin zone, $q<0.16$ r.l.u., is due to the Coulomb interaction, the polarizability, or the combination of the two.}

Physically, the polarizability $\Pi(\mathbf{q},\omega)$ can be understood as the density response of the system with a completely screened Coulomb interaction (i.e. the system with \textit{neutral} rather than charged density fluctuations) \cite{mahan,nozieres99}. The polarizability $\Pi(\mathbf{q},\omega)$ and susceptibility $\chi(\mathbf{q},\omega)$ are related by \cite{mahan,nozieres99},

\begin{equation}
\chi(\mathbf{q},\omega) = \frac{\Pi(\mathbf{q},\omega)}{\epsilon_{\infty}-V(\mathbf{q})\Pi(\mathbf{q},\omega)},
\label{eqn:chi_to_pi}
\end{equation}

\noindent
where $V(\mathbf{q})$ is the Coulomb interaction and $\epsilon_\infty$ is the background dielectric constant, which for Bi-2212 is $\epsilon_\infty = 4.5$ \cite{levallois16}.

In practice, the most important difference between the polarizability and the susceptibility is that the former does not feature plasmon excitations, whose existence is a consequence of the direct Coulomb interaction between particle and hole excitations. Instead, the polarizability $\Pi(\mathbf{q},\omega)$ reveals the particle-hole excitation spectrum itself.

We argued in Ref. \cite{mitrano18} that $\Pi(\mathbf{q},\omega)$ could be extracted from the M-EELS data by assuming a two-dimensional form for the Coulomb interaction, 

\begin{equation}
V(\mathbf{q}) \propto \frac{e^{- q d}}{q}.
\label{eqn:V_approx}
\end{equation}

\noindent
where $d$ is on the order of the interlayer spacing in Bi-2212. Here, we apply this procedure to all four doping values by extrapolating $\chi''(\mathbf{q},\omega)$ with a $1/\omega^2$ tail and Kramers-Kronig transforming to acquire the real part, $\chi'(\mathbf{q},\omega)$. We then evaluate $\Pi(\mathbf{q},\omega)$ for each doping from Eq. \ref{eqn:chi_to_pi} using $d=15.62 \,\textrm{\AA}$, which is close to the bilayer spacing for Bi-2212, and a proportionality constant in Eq. \ref{eqn:V_approx} for $V(\mathbf{q})$ of $7.9 \cdot 10^3 \,\textrm{eV}\cdot\textrm{\AA}^2$ (see Appendix A).  The result is displayed in Fig. 2, which shows the scaled quantity, $\Pi''(\mathbf{q},\omega)/q^2$, at low temperature for each of the four doping values. The spectra for all momenta collapse onto a single curve that is different for each doping value. This collapse implies that $\Pi(\mathbf{q},\omega)$ is constant for all momenta (again, apart from an overall $q^2$ factor required by the sum rule), even down to the lowest momentum measured, $q=0.1$ r.l.u. 

The overarching point of Fig. 2 is that particle-hole excitations in Bi-2212 cannot propagate, at least for energies $\omega > 0.1$ eV. The modest $q$-dependence in the M-EELS data below 0.16 r.l.u., where the plasmon-like maximum is visible in the data, is purely an effect of $V(\mathbf{q})$ in Eq. 2. The particle-hole spectrum itself, $\Pi''(\mathbf{q},\omega)$, appears to be constant even down to the lowest momentum studied, $q=0.1$ r.l.u. This is extremely surprising, since the individual quasiparticles in this energy range exhibit significant dispersion \cite{damascelli03,kaminski03,vishik2010,chatterjee2011}. We conclude that the textbook Lindhard description of the polarizability, in terms of particle-hole excitations across a Fermi surface and RPA treatment of the Coulomb interaction \cite{mahan,nozieres99}, completely fails in Bi-2212. This failure occurs at all compositions we examined---even overdoped materials that are usually considered to be good Fermi liquids \cite{keimer15}.

A second implication of Fig. 2 is that the peak in M-EELS data at low temperature in the overdoped regime (Fig. 1(e)-(h)) is not a plasmon. While the overdoped susceptibility in Fig. 1(e)-(h) exhibits a peak in a similar energy range as Fig. 1(a) with a local maximum near $\sim 0.8 - 1$ eV, the peak in the overdoped susceptiblity in Fig. 1(e)-(h) is also present in the polarizability (Fig. 2(d)), which as discussed above does not show plasmon effects. The peak in $\chi''(\mathbf{q},\omega)$ in the low-temperature overdoped regime is thus inherited from $\Pi''(\mathbf{q},\omega)$ and it is therefore more appropriate to think of it as a feature of the particle-hole excitation spectrum rather than as a collective, plasmon mode.

A final, striking implication of Fig. 2 is that, at energy $\omega > 0.1$ eV, the polarizability factors, i.e., $\Pi''(\mathbf{q},\omega) \sim f(\mathbf{q}) \cdot g(\omega)$, where $f(\mathbf{q}) \sim q^2$. Since the original literature on the marginal Fermi liquid (MFL) phenomenology of the cuprates \cite{varma2015,varma17}, this factoring has been considered a a key signature of ``local quantum criticality," an exotic phase of matter in which the spatial correlation length $\xi_x \sim \ln \xi_t$, where $\xi_t$ is the temporal correlation length \cite{varma89,varma2015,varma17,mitrano18,krikun2018}. One can think of such a phase as being characterized by a dynamical critical exponent $z=\infty$. The new lesson we have learned here is that this factoring takes place at all measured dopings and temperatures in the phase diagram. It is therefore a general materials property of Bi-2212, and not a feature of a critical point or a particular doping value.

\begin{figure*}[t]
	\includegraphics[width=\textwidth]{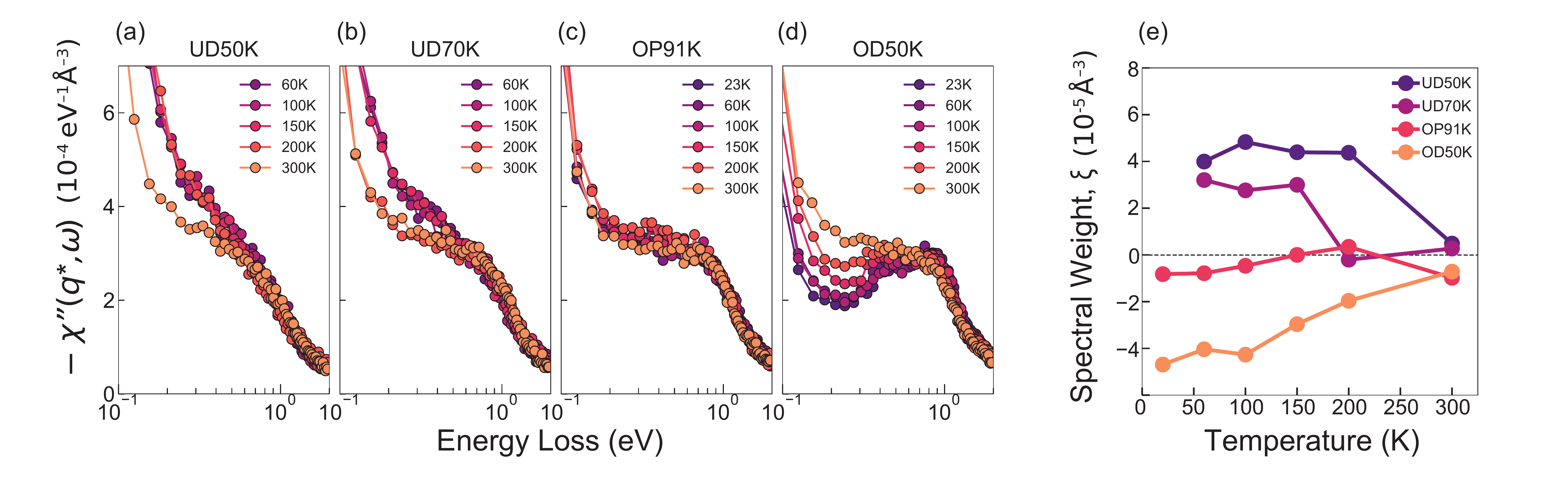}
	\caption{(a)-(d) Temperature dependence of $\chi^{\prime\prime}(q^*=0.24,\omega)$ for each doping level studied. Note that the energy axis is plotted on a log-scale to emphasize the spectral weight change below 0.5 eV. (e) Doping and temperature dependence of the spectral weight change $\xi$, as defined in the main text to be the \textcolor{black}{spectral weight} between 0.1 eV to 0.5 eV referenced to optimal doping at 150 K. The crossover from spectral weight accumulation to depletion with doping is evident, as is the sign-reversal at optimal doping.}
	\label{tempdep}
\end{figure*}

\textcolor{black}{\section{Spectral weight trend}}

The changes in spectral weight follow a distinct trend across the phase diagram. The fine temperature dependence is displayed in Fig. 3, which shows spectra at a fixed momentum $q^*=0.24$ r.l.u. at which $\chi''(\mathbf{q},\omega) \approx \Pi''(\mathbf{q},\omega)$. At optimal doping the spectra are temperature-independent, but the overdoped material shows a suppression of spectral weight below 0.5 eV as the system is cooled, indicating the emergence of an energy scale \cite{mitrano18}. In the underdoped case, this trend reverses: the weight below 0.5 eV is {\it enhanced} as the system is cooled, exhibiting a power-law form at low temperature (Fig. 3(a)-(b)). This enhancement may be a consequence of slowing CDW fluctuations in underdoped materials \cite{dasilveneto2014,comin2016}. The optimally doped case corresponds to a turning point between regions with opposite behavior, where the resulting response is temperature-independent (Fig. 3(c)). The normal state at this doping corresponds to the center of the strange metal regime in which resistivity is linear over the widest temperature range \cite{chatterjee2011,keimer15}.

A distinct trend is now clear: The M-EELS response is featureless and doping-independent at room temperature, but not at low temperature (Fig. 1). At the same time, the response is temperature-independent at optimal doping, but not at other dopings (Fig. 3(a)-(d)). The overall behavior may be summarized using a parameter that quantifies the deviation of the response from its featureless form at high temperature,

\begin{equation}
\xi = -\int_{0.1 \textrm{eV}}^{0.5 \textrm{eV}} \left [ \chi''(q^*,\omega) - \chi''_{\textrm{ref}}(q^*,\omega) \right ] d\omega,
\end{equation}

\noindent
 where the reference spectrum, $\chi''_{\textrm{ref}}(q^*,\omega)$, is taken at optimal doping at $T=150$ K and $q^*=0.24$ r.l.u. The quantity $\xi$ measures, as a function of temperature and doping, the degree to which the spectra deviate from the integrated spectral weight of $\chi''_{\textrm{ref}}(q^*,\omega)$ which has a constant value of $12.94\cdot 10^{-5} \,\textrm{\AA}^{-3}$. Note that this parameter is also a measure of the change in the Coulomb energy of the system \cite{levallois16}.

The behavior of $\xi$ is summarized in Fig. 3(e). Its value is small at 300 K for all four dopings. As the system is cooled, $\xi$ becomes negative in the overdoped case, where spectral weight is suppressed at low temperature, and positive for underdoped materials, where the weight is enhanced. At optimal doping, where the response is always featureless, $\xi$ remains small at all temperatures. Note that the value of $\xi$ as a function of doping at low temperature shows the same change in sign as the Coulomb energy determined from ellipsometry experiments \cite{levallois16}, though the magnitude of the effect observed here is significantly larger (Appendix B).

\begin{figure}[h]
	\includegraphics[width=8.4cm]{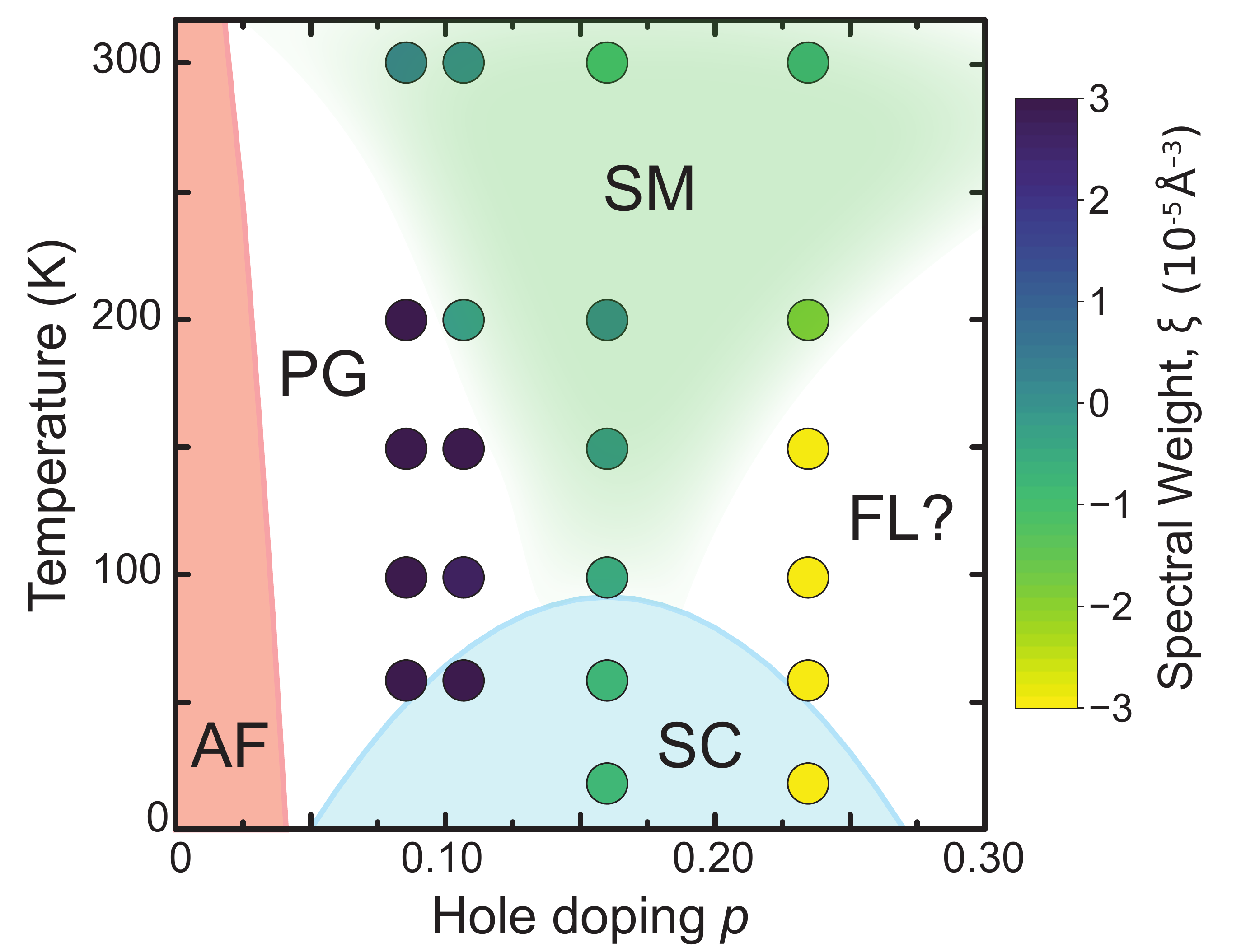}
	\caption{Evolution of the M-EELS continuum spectral weight across the Bi-2212 phase diagram, which was constructed from the data in Refs. \cite{vishik2010,chatterjee2011,barisic2013,proust2016}. Here AF=antiferromagnet, SC=superconductor, PG=pseudogap, FL=Fermi liquid, SM=strange metal. 
The colored points represent the spectral weight change, $\xi$, in Fig. 3(e). $\xi$ is small throughout the SM region in which the continuum remains featureless. Outside the SM, the continuum spectral weight changes rapidly with a different sign in the underdoped and overdoped regimes.
}
	\label{phasediagram}
\end{figure}

The trend is illustrated in another way in Fig. 4, which shows the spectral weight change $\xi$ as a function of doping and temperature superposed on the known phase diagram of Bi-2212 constructed from the phase boundaries in Refs. \cite{vishik2010,chatterjee2011,barisic2013,proust2016}. The region over which $\xi$ is small, in which the response remains featureless, closely coincides with the fan-shaped, strange metal region. Fig. 4 suggests a connection between a featureless form for the density response and the existence of a linear-in-$T$ normal state resistivity.

\section{Discussion}
The highly unconventional behavior of the density response across the Bi-2212 phase diagram can be summarized as consisting of a flat continuum of density fluctuations within the fan-like, strange metal region, and dramatic changes in spectral weight up to 0.5 eV outside this region with a sign-reversal at optimal doping. Here, we discuss the implications of the presence of this continuum for some other widely known properties of the cuprates. 

First, the continuum provides a natural, qualitative explanation for the normal state quasiparticle lifetimes in Bi-2212 \cite{damascelli03,kaminski03,vishik2010,chatterjee2011}. Broadly speaking, the poorest quasiparticle coherence is observed in underdoped materials at low temperature. The coherence increases with increasing doping, with the longest lifetimes observed at low temperature on the overdoped side \cite{vishik2010,chatterjee2011}. This behavior can be understood by recognizing that the continuum we observe should provide a decay path for quasiparticle damping, amplifying the imaginary part of the self-energy, $\Sigma''(\omega)$ \cite{damascelli03,vishik2010,chatterjee2011}. The strong damping of quasiparticles in underdoped materials can then be understood as a consequence of the enhancement of the M-EELS continuum at low temperature in this regime (Fig. 3(a)-(b)). Similarly, the increased quasiparticle coherence in overdoped materials arises because the continuum in this regime is suppressed (Fig. 3(d)).

At optimal doping, in the strange metal phase, the imaginary part of the self-energy, $\Sigma''(\omega)$, is linear in $\omega$, which has been shown to be consistent with an Eliashberg function that is frequency-independent \cite{bok16}. This behavior is highly consistent with the observation that the continuum is frequency-independent in this regime (Fig. 3(c)), and suggests that the M-EELS data may have a direct relationship with the Eliashberg function itself.

It is critical to keep in mind, however, that while the continuum provides a clear decay channel for quasiparticles, the mechanism by which the quasiparticles in turn generate the continuum is less clear. Because the quasiparticles are highly dispersive in Bi-2212 \cite{chatterjee2011,vishik2010}, a standard Lindhard calculation of the density response using RPA \cite{nozieres99,mahan} would yield a continuum that is also highly momentum-dependent, which is inconsistent with the M-EELS data at all compositions studied. It may be the case that beyond-RPA effects, such as excitonic, local field, or vertex correction effects dominate the response properties of strange metals. New theoretical approaches using nonperturbative techniques going beyond RPA, such as those based on the AdS-CFT correspondence, may provide progress in understanding the density response of the strange metal \cite{krikun2018,zaanen2019}.

It is worth considering whether the momentum-independence of the density fluctuations could be a consequence of strong disorder, which might eliminate momentum conservation by explicitly breaking translational symmetry, resulting in $q$-integrated response functions in all measurements. Even in materials in which the degree of disorder is low by traditional crystallographic standards, translational symmetry could be broken by emergent electronic heterogeneity, for which there is ample evidence in the cuprates \cite{fujita2012}. Such heterogeneity has even been proposed as the origin of the linear-in-$T$ resistivity in strange metals \cite{pelc2019}.
Unfortunately, such a view of the M-EELS data is inconsistent with ARPES and STM studies, which report clearly dispersing quasiparticle excitations in Bi-2212 where the charge response is $q$-independent \cite{damascelli03,reber15,bok16,vishik2010,chatterjee2011,fujita2012}. Moreover, recent RIXS experiments on electron-doped cuprates, which are not strange metals but have a similar degree of disorder \cite{armitage2010}, have clearly shown conventional dispersing plasmon excitations, in agreement with Hubbard model-based RPA calculations \cite{hepting2018}. Also, dispersing collective modes have been observed with M-EELS in other materials with similar degree of disorder \cite{kogar2017}. So it seems unlikely that disorder is the {\it sole} cause of the $q$-independence we see. On the other hand, it remains possible that the strong coupling physics of Bi-2212 conspires with disorder in such as way as to make momentum irrelevant in the density response, though not other observables.

Finally, we consider the question of whether the continuum could be a sign of some kind of quantum critical behavior. In this view, the fan-like structure implied by Fig. 4 might indicate a crossover near a doping $p_c  \sim 0.16$ that could be identified as a quantum critical point (QCP), as suggested by many authors \cite{sachdev2010,keimer15,badoux2016,damascelli03,kaminski03,ando04,hussey08,sterpetti17}. 
This interpretation is problematic, since few expected signatures of quantum criticality are present in the density fluctuation spectra. For example, no soft collective mode, with energy falling to zero at $p_c$, is visible in the data. A spectral weight rearrangement is observed away from $p_c$ in both underdoped and overdoped samples, however its energy scale $\sim 0.5$ eV is more than an order of magnitude larger than the crossover temperature, $\sim 200$ K, over which this rearrangement takes place (Fig. 3(a)-(d)). Moreover, the response functions near $p_c$ do not exhibit any momentum dependence, which is expected in the usual Hertz-Millis picture of a quantum phase transition \cite{sachdev99}. The factoring of $\Pi''(\mathbf{q},\omega)$ we observe (Fig. 2) has been cited as evidence for local criticality, which has been argued to be a feature of an exotic QCP \cite{varma2015,varma17}. But this factoring is observed everywhere in the phase diagram, not just in the vicinity of $p_c$. Whether some kind of exotic critical point might explain the peculiar physics taking place remains, for now, an open question.

In summary, we have shown that Bi-2212, despite being a good conductor by most standards, does not exhibit well-defined plasmon excitations that are dispersing and long-lived anywhere in its phase diagram. Instead, this material exhibits a featureless, momentum-independent continuum in the density response throughout the fan-shaped, strange metal region of the phase diagram. Outside this fan, the response undergoes dramatic changes in spectral weight up to 0.5 eV exhibiting a sign-reversal at optimal doping. Our study establishes a featureless continuum as a defining property of the mysterious strange metal phase in Bi-2212 and places it at a crossover between two regimes with opposite trends in their charge susceptibility. Major open questions remain concerning the relationship between the density response and models based on quasiparticle scattering, disorder, or quantum critical fluctuations. A new kind of theory of interacting matter may be needed to explain the existence of this phase and its connection to other exotic phenomena such as high temperature superconductivity.

\section{Acknowledgments}
We thank D. van der Marel, P. W. Phillips, C. M. Varma, N. D. Goldenfeld, A. J. Leggett, W. E. Pickett, and J. Zaanen for enlightening discussions, M. Tran for supplying the ellipsometry data used to evaluate the $f$-sum rule, and I. El Baggari for useful sample preparation advice. This work was supported by the U.S. Department of Energy, Office of Basic Energy Sciences grant no. DEFG02-06ER46285. P.A. gratefully acknowledges support from the EPiQS program of the Gordon and Betty Moore Foundation, grant GBMF4542. Crystal growth was supported by DOE grant DE-SC0012704. B.U. acknowledges NSF CAREER grant DMR-1352604. M.M. acknowledges  support  by  the  Alexander  von  Humboldt Foundation through the Feodor Lynen Fellowship program. C.D. acknowledges use of the electron microscopy facilities in the Eyring Materials Center at Arizona State University.

\appendix
\section{APPENDIX A: DETERMINING $\Pi(\mathbf{q},\omega)$}

At large momentum, the Coulomb interaction $V(\mathbf{q})$ is small so the susceptibility, $\chi(\mathbf{q},\omega)$, and the polarizability, $\Pi(\mathbf{q},\omega)$, are nearly the same (Eq. 2). At small momenta, particularly $q < 0.16$ r.l.u., relating them requires knowledge of $V(\mathbf{q})$. The precise functional form of $V(\mathbf{q})$ can depend on the material geometry, especially in quasi-2D systems \cite{reed2010} or at the surface of a layered material. Understanding this relation for M-EELS is still a work in progress. Nevertheless, as argued in Ref. \cite{mitrano18}, the approximate form shown in Eq. \ref{eqn:V_approx} is a sensible phenomenological starting point as it exhibits the same functional behavior as that of an infinite (i.e. not surface terminated) layered system \cite{reed2010}. 

To determine the proportionality constant in Eq. 3, we follow a similar procedure to Ref. \cite{mitrano18}. Noticing that in the limit of large momentum transfer, $\chi(\mathbf{q},\omega) \approx \Pi(\mathbf{q},\omega) / \epsilon_{\infty}$, we treat $V$ as a fit parameter, $V_{\textrm{fit}}$, and determine the values of $V_{\textrm{fit}}$ where $\Pi$ most closely resembles its value at the highest momentum measured, $q=0.5 \,\textrm{r.l.u.}$, by minimizing the function,

\begin{equation}
\frac{1}{\epsilon_{\infty}}\vert \Pi(\mathbf{q},\omega) - \Pi(0.5,\omega) \vert = \vert \frac{\chi(\mathbf{q},\omega)}{1+V_{\textrm{fit}}(q) \chi(\mathbf{q},\omega)} - \chi(0.5,\omega) \vert.
\label{eqn:Vfit}
\end{equation}

\begin{figure}[h]
	\includegraphics[width=8.4cm]{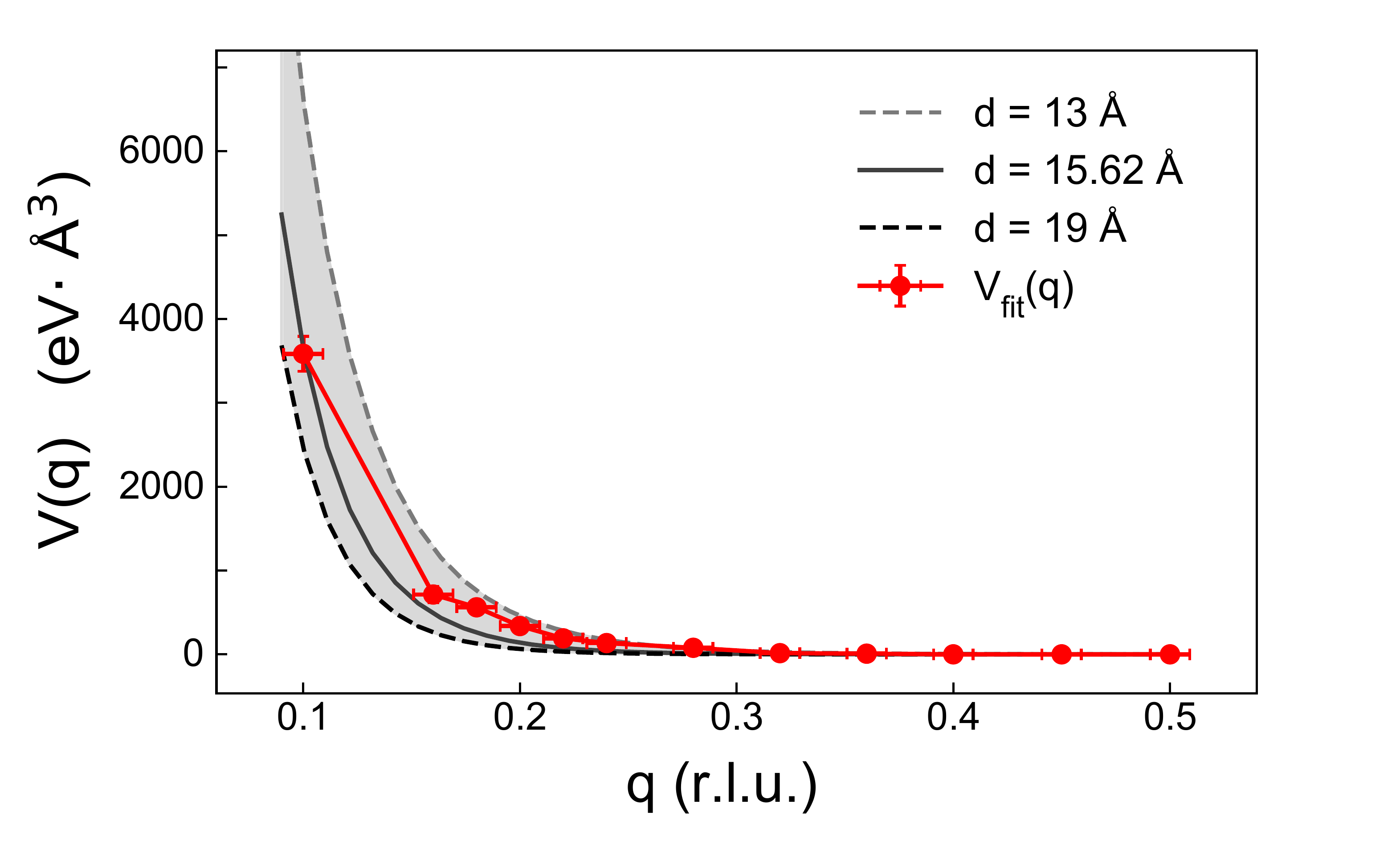}
	\caption{$V_{\textrm{fit}}(q)$ determined by minimizing Eq. \ref{eqn:Vfit} for OD50K, and the subsequent fit using Eq. \ref{eqn:V_approx}, obtaining a proportionality constant of $(7.9 \pm 0.3) \cdot 10^3 \,\textrm{eV}\cdot\textrm{\AA}^2$. Vertical error bars reflect statistical fitting uncertainties, while horizontal error-bars reflect the momentum resolution of the M-EELS instrument of $0.03 \,\textrm{\AA}^{-1}$. For reference, curves evaluated with $d=13 \,\textrm{\AA}$ and $d=19 \,\textrm{\AA}$ are also shown. }
	\label{Vprop_constant}
\end{figure}

After obtaining $V_{\textrm{fit}}(q)$, we fit to it with the functional form of Eq. 3 using $d=15.62 \textrm{ \AA}$ to determine the proportionality constant. The fits are shown in Fig. \ref{Vprop_constant}, along with curves with alternate values of $d$ shown for reference. The fitted proportionality constant was found to be $(7.9 \pm 0.3) \cdot 10^3 \,\textrm{eV}\cdot\textrm{\AA}^2$. Considering a rough estimate for the proportionality constant of $4\pi e^2 d = 2.83 \cdot 10^3 \,\textrm{eV}\cdot\textrm{\AA}^2$ is within a factor 3, the fit value is quite reasonable given the systematic uncertainties.

\section{APPENDIX B: CONSISTENCY BETWEEN M-EELS AND OTHER ELECTROMAGNETIC PROBES}

Our M-EELS data are consistent, in most respects, with previous studies of the electromagnetic properties of cuprates. Like M-EELS \cite{vig17}, spectroscopic ellipsometry \cite{bozovic1990,levallois16} and transmission EELS experiments with modest energy resolution (i.e., elastic linewidths of $\geq$ 0.3 eV) \cite{wang1990,nucker1989,nucker1991} also observe an overdamped plasmon excitation at $\sim$ 1 eV energy. At a small in-plane momentum, $q=0.05$ r.l.u., M-EELS and transmission EELS give nearly identical plasmon energy and lineshape \cite{vig17}. Further, the plasmon energy measured with ellipsometry is independent of doping \cite{terasaki90,uchida91,liu99,levallois16}, starkly contrasting with the behavior of a normal conductor \cite{nozieres99,mahan,kogar2015}, suggesting it may arise from a continuum of the sort we see. 

Still, it is clear something qualitative must change in the charge response as $q \rightarrow 0$. For one, the plasmon lineshape observed with ellipsometry, which is strictly a $q=0$ probe, is 50\% narrower than the M-EELS and transmission EELS data at the lowest momentum studied \cite{vig17}. Further, the spectral weight in ellipsometry grows with decreasing temperature for all dopings by no larger than $\sim$3\% \cite{levallois16}, while the changes observed here are around 50\% and may have either sign (Fig. 3(e)). 

More fundamentally, the factoring of the polarizability, $\Pi''(q,\omega) = f(q) \cdot g(\omega)$, cannot persist to zero momentum because it violates the compressibility sum rule \cite{mahan,nozieres99},

\begin{equation}
\lim_{q \rightarrow 0} \Pi(q,0) = -n^2 \kappa,
\end{equation}

\noindent
where $\kappa = - ({\partial V}/{\partial P})_N / V$ is the compressibility. By definition, $\kappa=0$ for an insulator while for a metal $\kappa$ is a constant. If the factoring of $\Pi$ persisted all the way to $q \rightarrow 0$ the $f$-sum rule (Eq. \ref{eqn:chi_to_pi}) would require $\Pi(q,0) \sim q^2$ in the limit of small momentum, implying that the system is an insulator. However Bi-2212 is a metal, of course, meaning this factoring cannot persist as $q \rightarrow 0$. Full reconciliation between $q=0$ probes like ellipsometry and finite-$q$ techniques may require new theoretical ideas about non-commutativity of limits \cite{setty2018}. Encouragingly, however, our $\xi$, which represents integrated spectral weight change, exhibits the same trend with doping as the Coulomb energy reported in Ref. \cite{levallois16}, changing sign at optimal doping in both studies. Forthcoming studies with higher momentum resolution will determine the nature of the crossover region between the two techniques.

We note that, at larger momenta, transmission EELS studies of Bi-2212 by different groups report conflicting results. Using 60 keV electrons in an EELS spectrometer employing Wien filters and a $30 \,\textrm{meV}$ quasi-elastic line, Terauchi et al. reported a featureless continuum with a 1 eV cutoff energy similar to what we report here \cite{terauchi1999}. On the other hand, using 170 keV electrons with hemispherical analyzers and quasi-elastic line extending to $500 \,\textrm{meV}$, N\"ucker et al. observed no continuum at all, but instead report conventional plasmon behavior exhibiting normal,  Fermi liquid-like dispersion \cite{nucker1989,nucker1991}. More recently, transmission EELS studies by Schuster {\it et al.} of the underdoped cuprate Ca$_{2-x}$Na$_x$CuO$_2$Cl$_2$ also show evidence for a broad continuum, rather than conventional dispersing plasmons \cite{schuster2009}. 

\begin{figure}[h]
	\includegraphics[width=8.4cm]{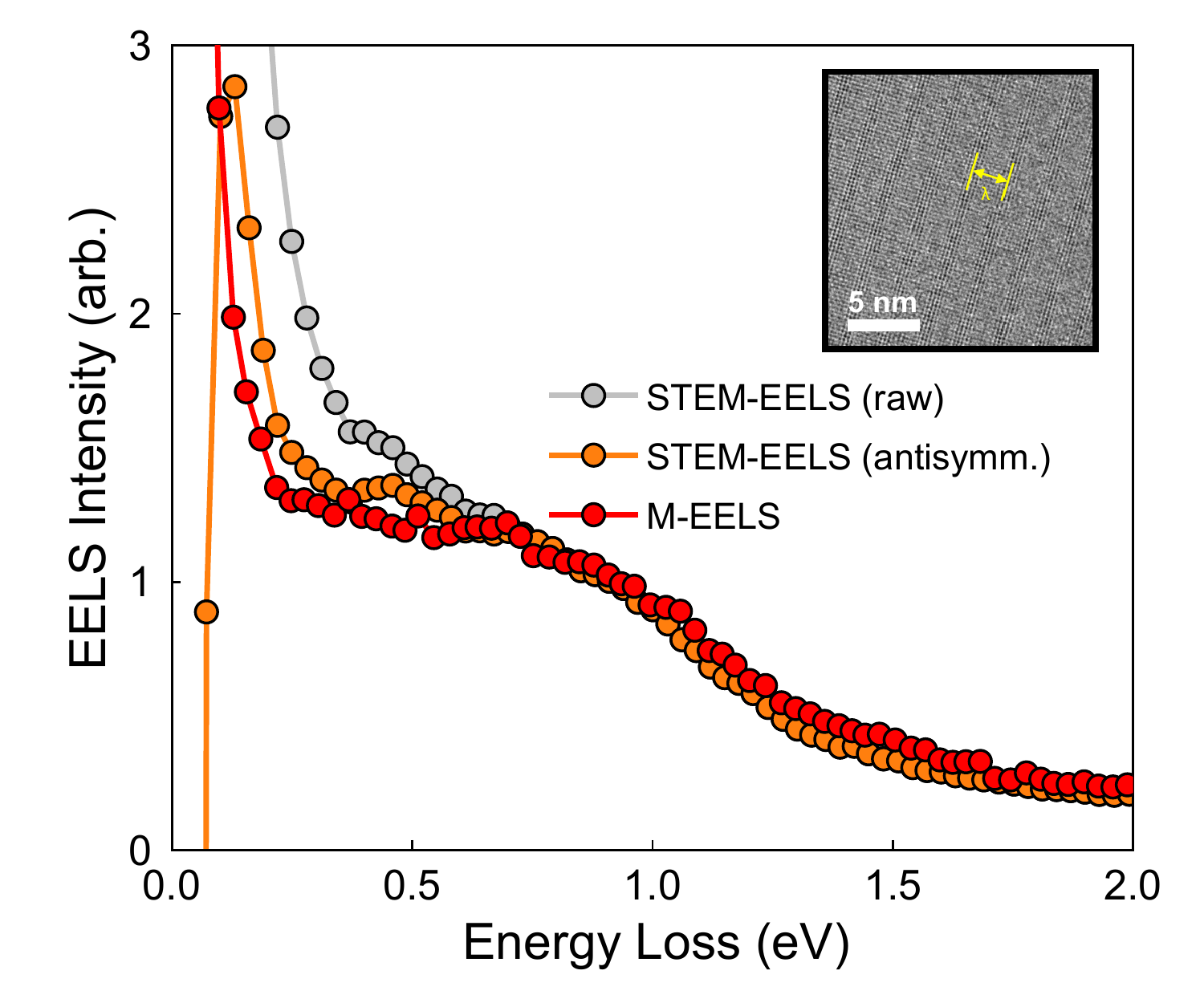}
	\caption{Comparison of M-EELS and Transmission STEM-EELS of OP91K Bi-2212 at 300 K. The raw STEM-EELS spectrum is shown in grey, and the anti-symmetrized spectrum is shown in orange. The M-EELS spectrum in red is taken at momentum transfer of $0.5$ r.l.u. (inset) Phase-contrast TEM image of the sample region investigated, showing the Bi-2212 supermodulation with wavelength $\lambda \approx 26 \mathrm{\AA}$, verifying the structural integrity of the sample used in STEM-EELS measurements.}
	\label{Nion_vs_MEELS}
\end{figure}

As an initial step towards establishing consistency between M-EELS and transmission EELS techniques, we have performed a preliminary measurement of OP91K Bi-2212 using a monochromated Nion UltraSTEM Scanning Transmission Electron Microscope EELS (STEM- EELS) with 10 meV resolution, as shown in Fig. \ref{Nion_vs_MEELS}. 
These measurements were performed at room-temperature with 60 keV electrons and convergence/acceptance semi-angles of 4 mrad. Samples were prepared using a ``powder" method, where a single crystal of OP91K Bi-2212 is crushed in a mortar and pestle and dispersed onto a holey carbon grid using ethanol. A flake was found with the correct orientation and estimated thickness of below 400 \AA. Sample crystallinity and ab-plane orientation was confirmed by verifying the well-known Bi-2212 supermodulations (see the inset of Figure \ref{Nion_vs_MEELS}). 
Note that the STEM-EELS measurements show precisely the same continuum with the same cutoff energy as M-EELS measurements. This indicates that the continuum is a bulk property of Bi-2212, and not a peculiarity of the surface.  

We note that, because STEM-EELS instruments are designed for high spatial-resolution at the expense of momentum-resolution, the STEM-EELS spectrum in Fig. 6 is momentum-integrated azimuthally in the ab-plane for $q\lesssim 0.63$ r.l.u. (i.e. approximately the first Brillouin Zone). Nevertheless, because M-EELS measurements have established that the strange metal continuum is largely momentum-independent, Fig. 6 represents a strong evidence that the continuum is a bulk effect. More detailed transmission EELS studies with both high energy- and momentum-resolution would serve as an even more stringent test.

\bibliography{Bi2212_UD_MFL}

\end{document}